\documentclass{ws-procs975x65}
\usepackage{amssymb,graphicx}

\def\l{{\cal L}}

\def\pd{\partial}

\def\be{\begin{equation}}
\def\ee{\end{equation}}
\def\bea{\begin{eqnarray}}
\def\eea{\end{eqnarray}}
\def\ie{\textit{i.e.} }
\def\etal{\textit{et.al.} }

\begin{document}

\title{Conformation changes and protein folding induced by $\phi^4$ interaction}

\author{M. Januar}
\address{Department of Physics, University of Indonesia,
Kampus UI Depok, Depok 16424, Indonesia}

\author{A. Sulaiman}
\address{Badan Pengkajian dan Penerapan Teknologi, BPPT Bld. II (19$^{\rm th}$
floor), Jl. M.H. Thamrin 8, Jakarta 10340, Indonesia\\
E-mail: asulaiman@webmail.bppt.go.id, sulaiman@teori.fisika.lipi.go.id}

\author{L.T. Handoko}
\address{Group for Theoretical and Computational Physics,
Research Center for Physics, Indonesian Institute of Sciences,
Kompleks Puspiptek Serpong, Tangerang, Indonesia\\
E-mail: handoko@teori.fisika.lipi.go.id, handoko@fisika.ui.ac.id,
laksana.tri.handoko@lipi.go.id}
\address{Department of Physics, University of Indonesia,
Kampus UI Depok, Depok 16424, Indonesia}

\begin{abstract}
A model to describe the mechanism of conformational dynamics
in protein based on matter interactions using lagrangian approach and imposing
certain symmetry breaking is proposed. Both conformation changes of proteins and
the injected
non-linear sources are represented by the bosonic lagrangian with an additional
$\phi^4$ interaction for the sources. In the model the spring tension of 
protein representing the internal hydrogen bonds is realized as the interactions
between  individual amino acids and  nonlinear sources. The folding
pathway is determined by the strength of nonlinear sources that propagate
through the protein backbone. It is also shown that the model reproduces the
results in some previous works.
\end{abstract}


\bodymatter

\section{Introduction}

The pathway of proteins are determined by the sequences of its amino acid
constituents. The time ordered of protein folding sequence leads from the
primary to the secondary and subsequent structures. The secondary structure
consists of the shape representing each segment of a polypeptide tied by
hydrogen bonds, van der Walls forces, electrostatic interaction and hydrophobic
effects. It is moreover formed around a group of amino acids considered as the
ground state. Then it is extended to include adjacent amino acids till the
blocking amino acids are reached, and the whole protein chain along the
polypeptide adopted its preferred secondary structure.

Our understanding on the underlying above-mentioned mechanism has unfortunately
not been at the satisfactory level. For instance, the studies based on
statistical analysis of identifying the probabilities of locating amino acids in
each secondary structure are still at the level of less than 75\% accuracy.
Moreover, the main mechanism responsible for a structured folding pathway have
not yet been identified at all. On the other hand, it is known that the protein
misfolding has been identified as the main cause of several diseases like
cancers and so on \cite{dobson}.

Recently, Mingaleev \etal have shown that the nonlinear excitations play an
important role in conformational dynamics by decreasing the effective
bending rigidity of a biopolymer chain leading to a buckling instability of the
chain \cite{mingaleev}. Following this understanding, a model to explain
the transition of a protein from a metastable to its ground conformation
induced by solitons  has been proposed \cite{jacob2}. In the model the mediator
of protein transition is the Davydov solitons propagating through the
protein backbone.

At present, the most reliable theoretical explanation for this kind of the
conformational dynamics of biomolecules is the so-called ab initio quantum
chemistry approach. This however requires astronomical computational power to
deal with realistic biological systems \cite{garcia, onuchi}. In contrary,
there are some phenomenological model describing the folding pathway as a
result of the interplay between the energy transfer from a solitary solution
that travels along the protein backbone and string tension \cite{berloff}.
There are also some attempts to describe the dynamics in term of elementary
biomatter using field theory approach \cite{sulaiman} and open quantum system
\cite{sulaiman2, sulaiman3}.

This paper follows the later approach, but starting from the first principle
using the lagrangian method to derive the responsible interactions and to
clarify its origins. The paper is organized as follows. First, the model and
the underlying assumptions are explained in detail. It is then followed by the
derivation of relevant equation of motions (EOMs). Summary and conclusion based
on the numerical analysis are given at the end of the paper.

\section{The models}

The model is an extension of the toy model proposed in \cite{jacob1}. More than considering a self-interaction mechanism as proposed in \cite{jacob1} and subsequently developed in \cite{jacob2,berloff}, more realistic model is introduced. In the model, the dynamics of amino acids forming proteins is initially considered as a free and linear system of bosonic matters. Further,  external nonlinear sources, like laser or light bunch, are introduced. The sources which propagate through the protein backbone interact each other with the amino acids to induce conformation changes. 

The model describes the conformation changes as the dynamics of amino acids
using a free and massive (relativistic) bosonic lagrangian as below,
\be
        \l_c = \left( \pd_\mu \phi_c \right)^\dagger  \left( \pd^\mu
\phi_c \right) + \frac{1}{2} m_{\phi_c}^2 \phi_c^\dagger \phi_c\; , 
        \label{eq:c}
\ee
where $\phi_c$ represents the conformation field. The hermite conjugate is 
$\phi^\dagger \equiv {(\phi^\ast)}^T$ for a general complex field $\phi$. On the
other hand, the nonlinear sources represented by the field $\phi_s$ are also
governed by a massless bosonic lagrangian,
\be
        \l_s = \left( \pd_\mu \phi_s \right)^\dagger  \left( \pd^\mu \phi_s
\right)  + V(\phi_s) \; , 
        \label{eq:s}
\ee
with an additional potential $V(\phi_s)$ taking the typical $\phi^4-$
self-interaction, 
\be
  V(\phi_s) = \frac{1}{4} \lambda \, (\phi_s^\dagger \phi_s)^2 \; ,
  \label{eq:vns}
\ee
where $\lambda$ is the coupling constant. It should be noted that both scalar
fields, $\phi_c = \phi_c(t,x)$ denotes the local curvature of the conformation
at position $x$ with $\phi_c(x) = 1$ or $0$ for $\alpha$ or  $\beta-$helix.

The choice of interactions in Eqs. (\ref{eq:c}) and (\ref{eq:s}) are justified
by the following considerations,
\begin{itemize}
 \item The conformation changes are assumed to be linear. It is actually not
necessarily massive. Although one can put by hand the mass term $m_{\phi_c}^2
\phi_c^\dagger
\phi_c$ in the lagrangian as written above, the massive conformational field
could also be generated dynamically through certain symmetry breaking as shown
later. 
 \item The source is assumed to be massless concerning the laser or light
source injected to the protein chains to induce the foldings.
\item Its non-linearity is realized by introducing the $\phi_s$
self-interaction which leads to the non-linear EOM.
 \item For the sake of simplicity, the lagrangian is imposed to be symmetry
under certain transformations, for instance in the present case is time
and parity  symmetry, \ie $\phi(t,x) \rightarrow -\phi(-t,-x)$ for
one-dimensional space.
\end{itemize}

We should remark here that the model is although written in a relativistic form,
after deriving relevant EOMs one can take its non-relativistic limits to obtain
final EOMs describing the desired dynamics. Secondly, instead of using the
vector electromagnetic field $A_\mu$ to represent the nonlinear sources, like
laser for instance, it is more convenient to consider the nonlinear source as a
bunch of light or laser such that one might represent it in a 'macrosocopic'
scalar field $\phi_s$.

Considering the dimensional counting and the invariance on time-parity
symmetry, the most general interaction between the conformation field and
nonlinear sources is,
\be
  \l_\mathrm{int} = -\Lambda \, (\phi_c^\dagger \phi_c) (\phi_s^\dagger \phi_s)
\; ,
\label{eq:int}
\ee
with $\Lambda$ denotes the strength of the interaction. Eqs. (\ref{eq:vns}) and
(\ref{eq:int}) lead to the total potential in the model,
\be
  V_\mathrm{tot} = \frac{1}{4} \lambda \, (\phi_s^\dagger \phi_s)^2 - \Lambda \,
(\phi_c^\dagger \phi_c) (\phi_s^\dagger \phi_s) \; .
  \label{eq:vt}
\ee
Eqs. (\ref{eq:c}), (\ref{eq:s}) and (\ref{eq:vt}) provide the underlying
interactions in the model.

Concerning the minima of the total potential in term of source field, that is
\be
  \left. \frac{\partial V_\mathrm{tot}}{\partial \phi_s} \right|_{\langle \phi_s
 \rangle, \langle \phi_c \rangle} = 0 \; ,
\ee
at the vacuum expectation values (VEV) of the fields yields the non-trivial
solution,
\be
  \langle \phi_s \rangle = \sqrt{\frac{2\Lambda}{\lambda}} \langle \phi_c
\rangle \; .
  \label{eq:vev}
\ee

Imposing certain local symmetry, namely the phase or U(1) symmetry to the above
total lagrangian, the VEV in Eq. (\ref{eq:vev}) obviously breaks the symmetry.
The symmetry breaking at the same time shifts the mass term for $\phi_c$
as follow,
\be
  m_{\phi_c}^2 \rightarrow \overline{m}_{\phi_c}^2 \equiv m_{\phi_c}^2 - \frac{2
\Lambda^2}{\lambda} \langle \phi_c \rangle^2 \; ,
\ee
from Eq. (\ref{eq:int}).

On the other hand, Eq. (\ref{eq:vev}) induces the 'tension force' which plays
an important role to enable folded pathway in the present model. This will be
discussed in the following section.

\section{EOMs and its behaviours}

Having the total lagrangian at hand, one can derive the EOM's using the
Euler-Lagrange equation,
\be
  \frac{\partial \l_\mathrm{tot}}{\partial \phi} - \partial_\mu \frac{\partial
\l_\mathrm{tot}}{\partial (\partial_\mu \phi)} = 0 \; ,
  \label{eq:eue}
\ee
where $\l_\mathrm{tot} = \l_c + \l_s + \l_\mathrm{int}$.

Substituting Eqs. (\ref{eq:c}), (\ref{eq:s}) and (\ref{eq:int}) into Eq.
(\ref{eq:eue}) in term of $\phi_c$ and $\phi_s$, one immediately obtains a set
of EOMs,
\bea
  \left( \frac{\partial^2}{\partial x^2} - \frac{1}{c^2}
\frac{\partial^2}{\partial t^2} - \frac{1}{\hbar^2} m_{\phi_c}^2 c^2 + 2
\Lambda \, \phi_s^2 \right) \phi_c & = & 0 \; , 
  \label{eq:eomc}\\
  \left( \frac{\partial^2}{\partial x^2} - \frac{1}{c^2}
\frac{\partial^2}{\partial t^2} + 2 \Lambda \, \phi_c^2 - 3 \lambda \, \phi_s^2 
\right) \phi_s & = & 0 \; .
  \label{eq:eoms}
\eea
Here the natural unit is restored to make the light velocity $c$ and $\hbar$
reappear in the equation. 

The last term in Eq. (\ref{eq:eoms}) determines the non-linearity of the EOM of
source. One should also put an attention in the last term of Eq.
(\ref{eq:eomc}), \ie $\sim k \, \phi_c$ with $k \sim 2 \Lambda \langle \phi_s 
\rangle^2$. This actually induces the tension force in the dynamics of
conformational field enabling the folded pathway as expected.

Hence, solving both EOMs in Eqs. (\ref{eq:eomc}) and (\ref{eq:eoms})
simultaneously would provide the contour of conformational changes in term of
time and one-dimensional space components.

\begin{figure}[t]
        \centering 
  \noindent
  \vspace*{1cm}\\
	\includegraphics[width=\textwidth]{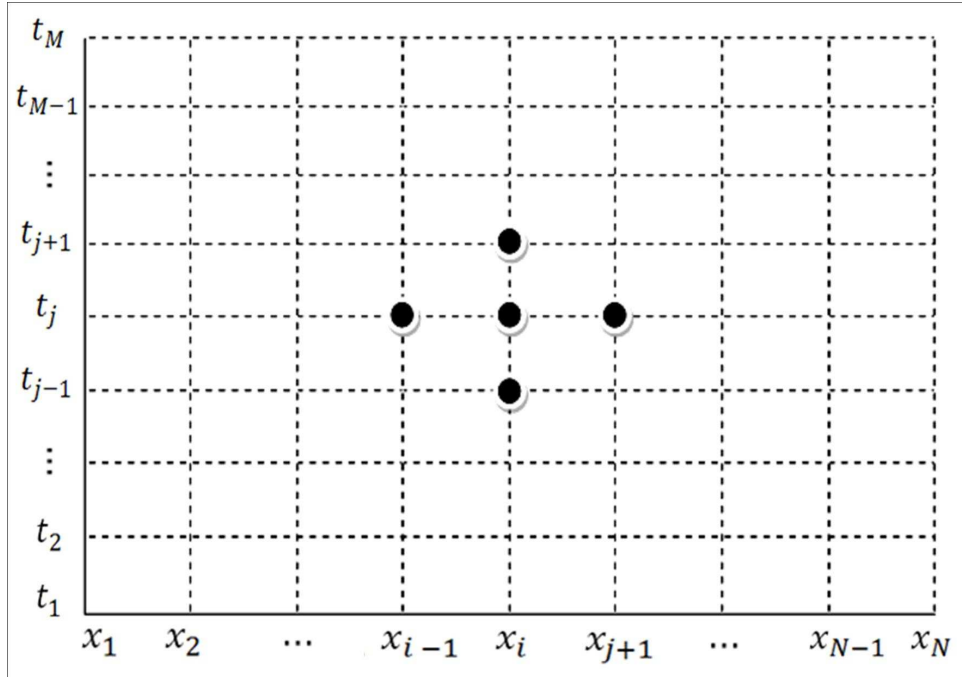}
        \caption{The discretized grid for solving the EOMs over the
coordinate space $R$.}
        \label{fig:1}
\end{figure}

\section{Numerical analysis}

Since the EOMs under consideration involves non-linear term, one should solve
them numerically. The numerical analysis and simulation in the present paper
are done using the finite difference method \cite{fink}. Throughout 
numerical works,  non-relativistic limit $v = {\partial x}/{\partial
t} \ll c$ and the following boundary conditions for both fields are deployed,
\be
\begin{array}{lcl}
 \phi_s(0,t) = \phi_s(L,t)=0\;\;\text{and}\;\;\phi_c(0,t)=\phi_c(L,
t)=0 & \text{for} & 0\le t \le b \; , \\
\phi_s(x,0)=f(x)\;\;\text{and} \;\;\phi_c(x,
0)=p(x) & \text{for} & 0\le x\le L \; , \\
\displaystyle \frac{\partial\phi_s(x,0)}{\partial t} =
g(x)\;\;\text{and}\;\;\frac{\partial\phi_c(x,0)}{\partial
t}=q(x) & \text{for} & 0 < x < L \; ,
\end{array}
\label{eq:bc}
\ee
with $f(x)$, $p(x)$, $g(x)$ and $q(x)$ are newly introduced auxiliary
functions. 
In finite difference scheme, it is more convenient to replace $\phi_s$ and 
$\phi_c$ with {\it u} and {\it w} respectively, and rewrite them in discrete
forms. Then, let us consider the coordinate space $R = \{(x,t):0\le x \le L,
0 \le t \le b\}$ discretized on a grid consisting of $(N-1) \times (M-1)$
rectangles with side length $\Delta x=\delta$ and $\Delta t=\epsilon$ shown in
Fig. \ref{fig:1}. Solving the equations over the grid gives us the desired
numerical solutions. 

Both coupled EOMs in Eqs. (\ref{eq:eomc}) and (\ref{eq:eoms}) are rewritten in
explicit discrete forms as follows,
\bea
u_{i,j+1} & = & 2u_{i,i}-u_{i,j-1}+c^2\epsilon^2
\left( \frac{u_{i+1,j}-2u_{i,j}+u_{i-1,j}}{\delta^2} +2\Lambda
w_{i,j}^2u_{i,j}-3\lambda u_{i,j}^3 \right) \; ,
\label{eq:u}\\
w_{i,j+1} & = & 2w_{i,i}-w_{i,j-1}+c^2\epsilon^2 
\left( \frac{w_{i+1,j}-2w_{i,j}+w_{i-1,j}}{ 
\delta^2}+2\Lambda u_{i,j}^2w_{i,j}
\right.\nonumber\\
&&\left. -\frac{c^2}{\hbar^2}m_{\phi_c}^2w_{i,j} \right) \; ,
\label{eq:w}
\eea
for $i = 2, 3, \cdots, N-1$ and $j = 2, 3, \cdots, M-1$. In order to calculate
all values of Eqs. (\ref{eq:u}) and (\ref{eq:w}), the initial values for two
lowest rows in Fig. \ref{fig:1} must be given. On the other hand, the value at
$t_1$ is fixed by the boundary conditions in Eq. (\ref{eq:bc}). The second
order of Taylor expansion can also be used to determine the values in
the second row. Therefore, the values at $\it t_2$ are determined by, 
\bea
u_{i,2} & = & f_i-\epsilon g_i+\frac{c^2\epsilon^2}{2} 
\left( \frac{f_{i+1}-2f_i+f_{i-1}}{\delta^2}+2\Lambda p_i^2 f_i - 3 \lambda
f_i^3 \right) \; , 
\label{eq:u2}\\
w_{i,2} & = & p_i-\epsilon q_i+\frac{c^2\epsilon^2}{2} 
\left( \frac{p_{i+1}-2p_i+p_{i-1}}{\delta^2}+2\Lambda
f_i^2p_i-\frac{c^2}{\hbar^2}m_{\phi_c}^2p_i \right) \; ,
\label{eq:w2}
\eea
for $i = 2, 3, \cdots, N-1$.

For the initial stage, suppose the nonlinear sources has a particular form 
$f(x)=2 \mathrm{sech}(2x) \, \mathrm{e}^{i2x}$ and $g(x)=1$
to generate the $\alpha$-helix, while $g(x) = q(x) = 0$ for the sake of
simplicity. Then, one can obtain the initial values in this case using Eqs.
(\ref{eq:u2}) and (\ref{eq:w2}). The subsequent values are
generated by substituting the preceeding values into Eqs. (\ref{eq:u}) and
(\ref{eq:w}). The higher order values can be obtained using iterative procedure.
  
\begin{figure}[t]
        \centering 
  \noindent
  \vspace*{1cm}\\
	\includegraphics[width=\textwidth]{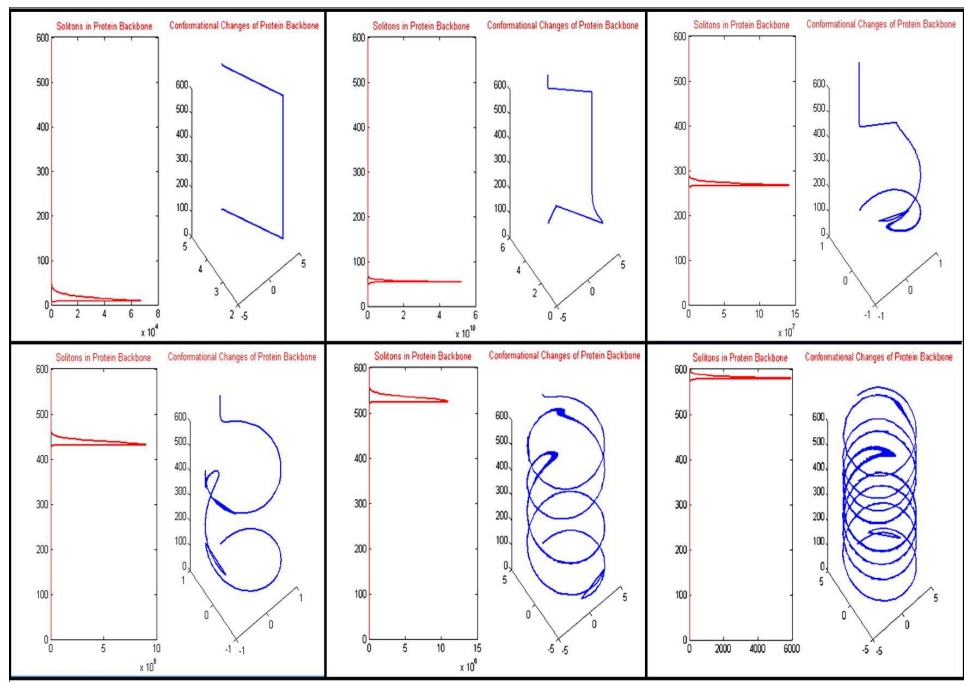}
        \caption{The soliton propagations and conformational changes on
the protein backbone inducing protein folding. The vertical axis in soliton
evolution denotes time in second, while the horizontal axis denotes its
amplitude. The conformational changes are on the $(x,y,z)$ plane.}
        \label{fig:2}
\end{figure}

The result is given in Fig. \ref{fig:2}. The left figure in each box describes
the propagation of nonlinear sources in protein backbone, while the right one
shows how the protein is folded. As can be seen in the figure, the
protein backbone is initially linear before the nonlinear source injection. As
the soliton started propagating over the backbone, the conformational changes
appear. It should be remarked that the result is obtained up to the second
order accuracy in Taylor expansion. In order to guarantee that the numerical
solutions do not contain large amount of truncation errors, the step sizes 
$\delta$ and $\epsilon$ are kept small enough. Nevertheless, this should be good
approximation to describe visually the mechanism of protein folding.

\section{Conclusion}

An extension of phenomenological model describing the conformational dynamics
of proteins is proposed. The model based on the matter interactions among the
relevant constituents, namely the conformational field and the nonlinear
sources represented as the bosonic fields $\phi_c$ and $\phi_s$. It has been
shown that from the relativistic bosonic lagrangian with $\phi_s^4$
self-interaction, the nonlinear and tension force terms appear naturally as
expected in some previous works \cite{berloff}. 

However, the present model has different contour since the EOMs governing the
whole dynamics are the linear and nonlinear Klein-Gordon equations. Note that
the original model by Berloff deployed the linear Klein-Gordon and nonlinear
Schrodinger equations.

Moreover, the present model has inhomogenous tension force, in contrast with
the homogeneous tension force in the Berloff's model, due to simultaneous
solutions of Eqs. (\ref{eq:eomc}) and (\ref{eq:eoms}). These lead to wrigling
folded pathways as shown in Fig. \ref{fig:2} which should be more natural than
the homogeneous one.

\section*{Acknowledgments}

The authors greatly appreciate fruitful discussion with T.P. Djun throughout the
work. AS thanks the Group for Theoretical and Computational Physics LIPI for
warm hospitality during the work. This work is partially funded by the Indonesia
Ministry of Research and Technology and the Riset Kompetitif LIPI in fiscal year
2010 under Contract no.  11.04/SK/KPPI/II/2010.

\bibliographystyle{ws-procs975x65}
\bibliography{folding}

\end{document}